\documentclass[conference]{IEEEtran}
\IEEEoverridecommandlockouts

\usepackage{cite}
\usepackage{amsmath,amssymb,amsfonts}
\usepackage{algorithmic}
\usepackage{graphicx}
\usepackage{textcomp}
\usepackage{xcolor}
\usepackage{csvsimple}
\usepackage{rotating}
\usepackage{subfig}
\usepackage{float}
\usepackage{url}
\usepackage[skip=8pt,font=scriptsize]{caption}

\usepackage{listings}
\usepackage{color}

\definecolor{dkgreen}{rgb}{0,0.6,0}
\definecolor{gray}{rgb}{0.5,0.5,0.5}
\definecolor{mauve}{rgb}{0.58,0,0.82}

\usepackage{wrapfig}
\usepackage{lscape}
\usepackage{rotating}
\usepackage{epstopdf}
\usepackage{array}
\usepackage{makecell}

\def\BibTeX{{\rm B\kern-.05em{\sc i\kern-.025em b}\kern-.08em
		T\kern-.1667em\lower.7ex\hbox{E}\kern-.125em}}
\graphicspath{ {fig/} }

\begin{document}
	
	\title{Simulation and Analysis of Distributed Wireless Sensor Network using Message Passing Interface }
	
	\author{
		\IEEEauthorblockN{Bhanuka Manesha Samarasekara Vitharana Gamage\IEEEauthorrefmark{1},
		Vishnu Monn Baskaran\IEEEauthorrefmark{2}}
		\IEEEauthorblockA{School of Information Technology,\\Monash University Malaysia,\\ Jalan Lagoon Selatan, 47500 Bandar Sunway,\\ Selangor Darul Ehsan, Malaysia.\\
		Email: \IEEEauthorrefmark{1}bsam0002@student.monash.edu,
		\IEEEauthorrefmark{2}vishnu.monn@monash.edu}
	}
	\maketitle
	
	\begin{abstract}
		Wireless Sensor Networks (WSN) are used by many industries from environment monitoring systems to NASA's space exploration programs, as it has allowed society to monitor and prevent problems before they occur with less cost and maintenance. This document aims to propose and analyze an efficient inter process communication (IPC) architecture using a nearest neighbor/grid based socket architecture. A parallelized version of the AES encryption algorithm is also used in order to increase the security of the WSN. First the proposed architecture is compared and contrasted against other well established architectures. Next, the benefits and drawbacks of the AES encryption algorithm is elucidated. The Message Parsing Interface (MPI) library in C is used for the communication while OpenMP is used for parallelizing the encryption algorithm. Next an analysis is performed on the results obtained from multiple simulations. Finally a conclusion is made that the grid based IPC architecture with AES parallel encryption helps WSNs maintain security in communication while being cost and power efficient to operate.

	\end{abstract}

	\begin{IEEEkeywords}
	wireless sensor networks, wsn, interprocess communication, ipc, grid based architecture, socket architecture, nearest neighbor architecture, advanced encryption standard, parallel aes, openmp, mpi, distributed computing
	\end{IEEEkeywords}
	
	\section{Introduction} \label{intro}
	
	The rapid growth of Internet of Things (IoT) devices and sensors has led to a high demand in efficient sensor architectures \cite{othman2012wireless}. Since sensors mainly run on embedded devices and battery powered systems, the need for power efficient architectures are on the rise \cite{guy2006wireless}. Currently, Wireless Sensor Network (WSN) architectures are the standard for low power sensor networks. To accomplish low power usage, each node in the WSN tries to reduce the amount of processing and communication done in them. In order to implement these WSNs, highly efficient Inter Process Communication (IPC) architectures are needed.
	
	An Inter Process Communication is a system that allows data to be communicated between two processes \cite{techopedia.com}. Pipes, Socket, File, Signal, Shared Memory and Message Queue are few of the different IPC architectures currently available \cite{onsman_2018}. Since, any given node has to communicate with either another node or the base station and the nodes can be located in different geographical locations, out of the given architectures, the socket architecture will be the most suitable, as it creates end points on each node to send and receive data. There are different design space architectures for socket based IPC architectures such as hyper cube, pyramid, tree and grid (nearest neighbor). But since a sensor node can only communicate with the four adjacent nodes i.e. nearest neighbors, a grid based socket IPC architecture would be the best method for the WSN. 
	
	One might argue that the pyramid based architecture will also be efficient as it converges the number of nodes as it comes to the top of the pyramid and each level of the pyramid can communicate with their adjacent nodes. But the main issue with this approach is that it is complex to build and maintain, and the number of messages needed to send a message from a bottom level to the top level is high. This also means that there is more chance of failure and if one middle layer node fails, then the whole sub pyramid of that node will also fail. Therefore, using a grid based nearest neighbor architecture allows us to ensure that no catastrophic failures in the architecture can happen as failure in one sensor node does not effect other nodes around it. The vulnerability of the system failing  when the base station fails, can be overcome by having a backup server or load balancer at the base station. Miller et al. \cite{miller1996parallel} provides a more in-depth analysis on the benefits and drawback of each architecture.
	
	Another major constraint of WSNs, is that not only do they have to operate on low power, but also be secure enough to broadcast sensitive sensor data. This is a major concern as most of the communication is done wirelessly and attackers can get the data easily \cite{wn}. To achieve this, state of the art efficient encryption standards are used by the system.
	
	In the following sections, we explore and analyze \textit{``How a grid based IPC architecture, that uses state of the art encryption standards can be made to run efficiently on a Wireless Sensor Network".} 
	
		\begin{figure*}[!t]
		\centering
		\caption{Dynamic Grid Layout}
		\subfloat[4 X 5 grid]
		{
			\centering
			\includegraphics[width=2.3in]{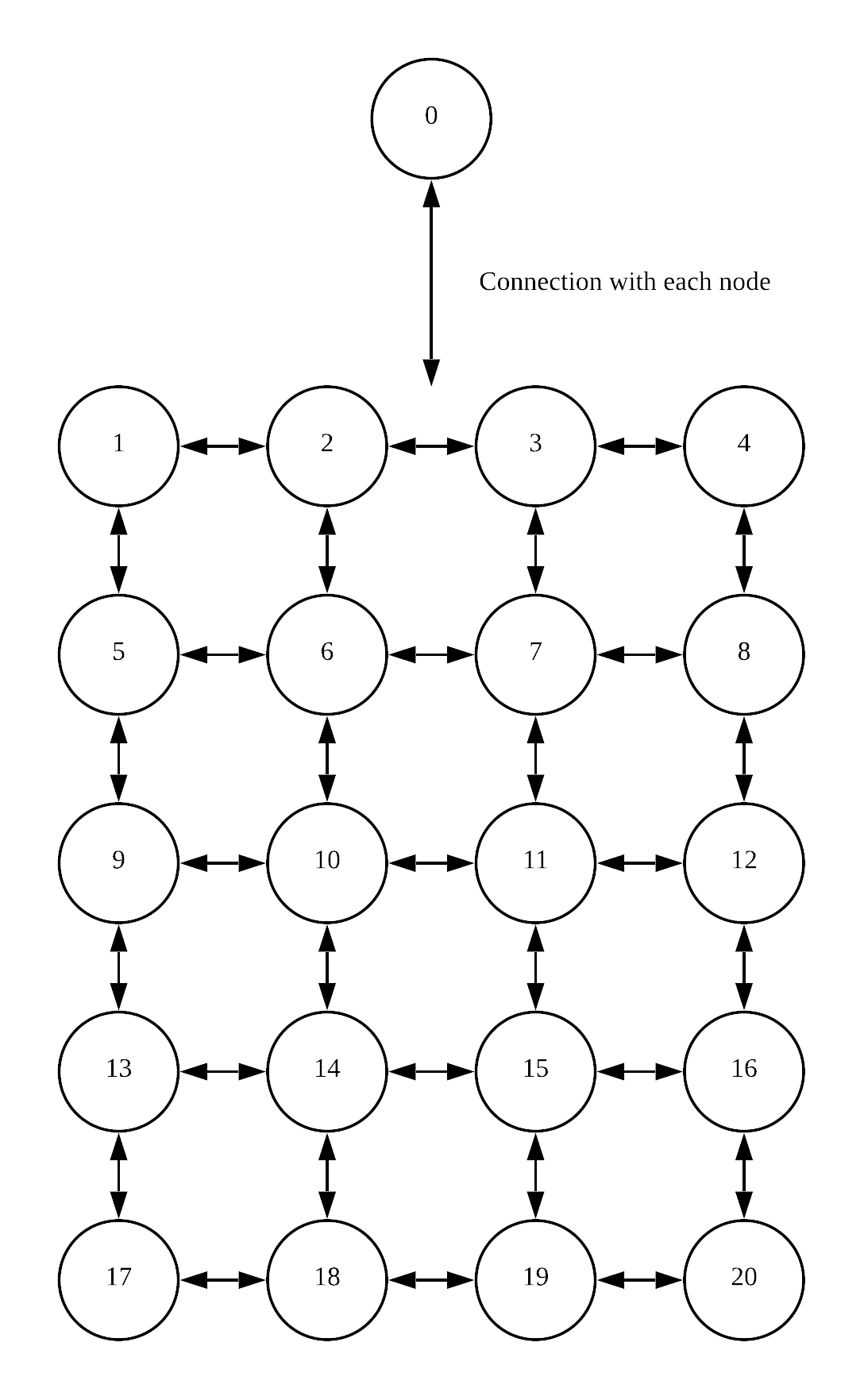}
			\label{gridlay1}
		}
		\\
		\subfloat[10 X 2 grid]
		{
			\centering
			\includegraphics[width=6in]{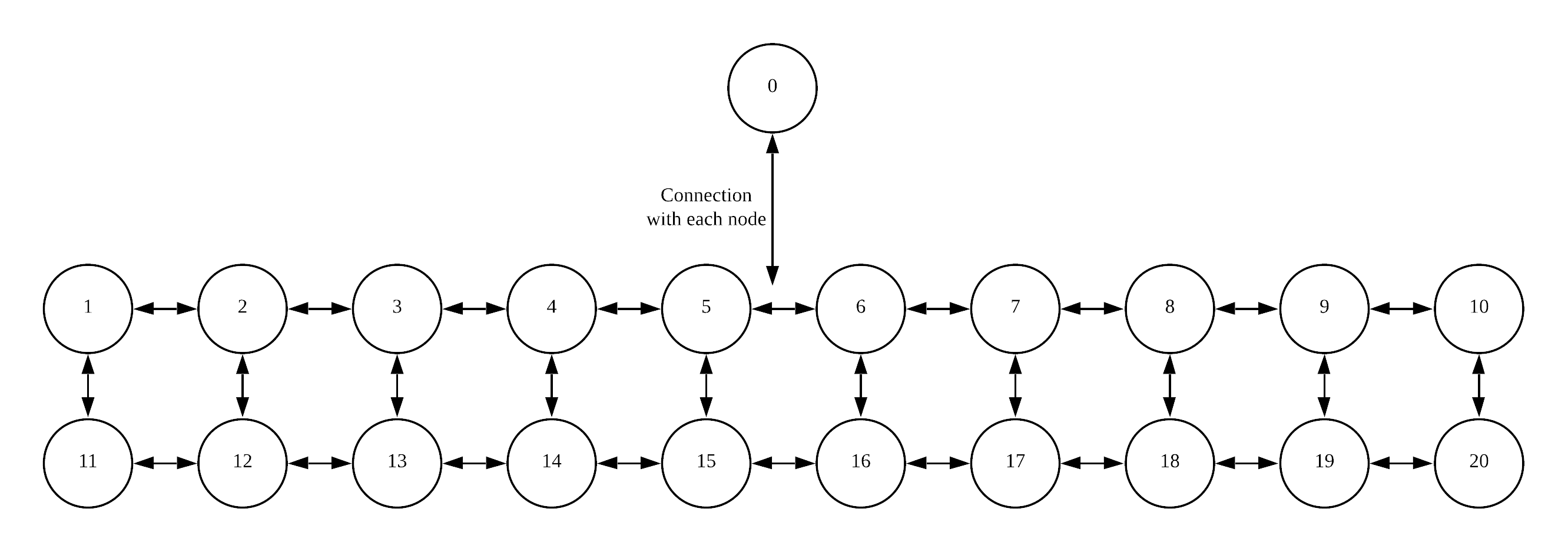}
			\label{gridlay2}
		}
		\label{gridlay}
	\end{figure*}

	\section{Theoretical Analysis and Inter Process Communication Design} \label{theoratical}
	
	The IPC Design used by the WSN is discussed in this section. The WSN uses a grid based architecture where each node is able to communicate with the adjacent node and the base station only.
	
	\subsection{ Inter Process Communication Grid Architecture}\label{IPCArc}
	
	The number of nodes used by the WSN depends on how many processes are allocated by the user at the start. If the user specifies the number of nodes as 21, then the system uses one node as the base and the rest as the sensor nodes. Using MPI, we allocate rank $0^{th}$ process to the base station and the rest of the nodes as the sensor nodes. The reason for selecting the base station as rank 0 is that only rank 0 get inputs from the user.
	
	The system uses a dynamic grid based architecture. So the layout of the nodes are based on the height and width specified by the user. Figure \ref{gridlay}  illustrates a simple example of how the user can rearrange a 21 node architecture. Figure \ref{gridlay1} illustrates how the grid is initialized when the width and height is 4 and 5 respectively, while Figure \ref{gridlay2} illustrates when its 10 and 2 respectively.
	
	After initializing, each node is able to communicate with the adjacent four nodes i.e left, right, top and bottom nodes and also with the base station. The base station is able to communicate with each node individually. The adjacent nodes of each of the sensor nodes are calculated using the width and height specified by the user. Equation \ref{row} and \ref{col} can be used to determine the row index and column index of a given node in the WSN. \emph{Note: Due to the base station being rank 0, we first minus 1 from the current rank}

	\begin{equation}
	row index = (rank - 1) \, // \, width
	\label{row}
	\end{equation}

	\begin{equation}
	column index = (rank - 1) \ mod \ width
	\label{col}
	\end{equation}
	
	After determining the row index and the column index, we can use Equations \ref{left}, \ref{right}, \ref{top} and \ref{bottom} to get the four adjacent nodes. 
	
	\begin{equation}
	left = row index \times width + column index
	\label{left}
	\end{equation}
	
	\begin{equation}
	right = row index\times width + column index + 2
	\label{right}
	\end{equation}
	
	\begin{equation}
	top = (row index - 1) \times width + column index + 1
	\label{top}
	\end{equation}

	\begin{equation}
	bottom = (row index + 1) \times width + column index + 1
	\label{bottom}
	\end{equation}
	
	\textit{Note :  The shift in the ranks due to the base station in rank 0 is accounted for in the equations and also a check needs to be done to ensure that the current rank is not an edge or corner node.} 
	
	Using this process, we do not need to send the rank values to the base station when an event occurs. The base station can use the rank of the incoming node to generate the four adjacent nodes in order. Therefore, this reduces the message size and optimizes the nodes to use less energy and the base to consume more energy for computation.
	
	Hoefler et al. \cite{hoefler2011scalable} mentions that the nearest neighbor architecture as one of the scalable process topologies. Therefore, this nearest neighbor algorithm used for WSN should be scalable as the number of sensors can increase with the growth in the field of IoT. Schneider and Hoefler then proposed methods to optimize Neighborhood Communications based on the 2D grid architecture we discussed above \cite{hoefler2012optimization}. Miller et al. \cite{miller1996parallel} also discussed about communication algorithms to be used with grid based architecture in broadcast and scatter/gather communication. Therefore, considering the prior research and the advantages it allows for WSNs, the 2D grid based nearest neighbor architecture will be used for our system.

	\begin{figure*}[!ht]
		\centering
		\includegraphics[width=6.5in]{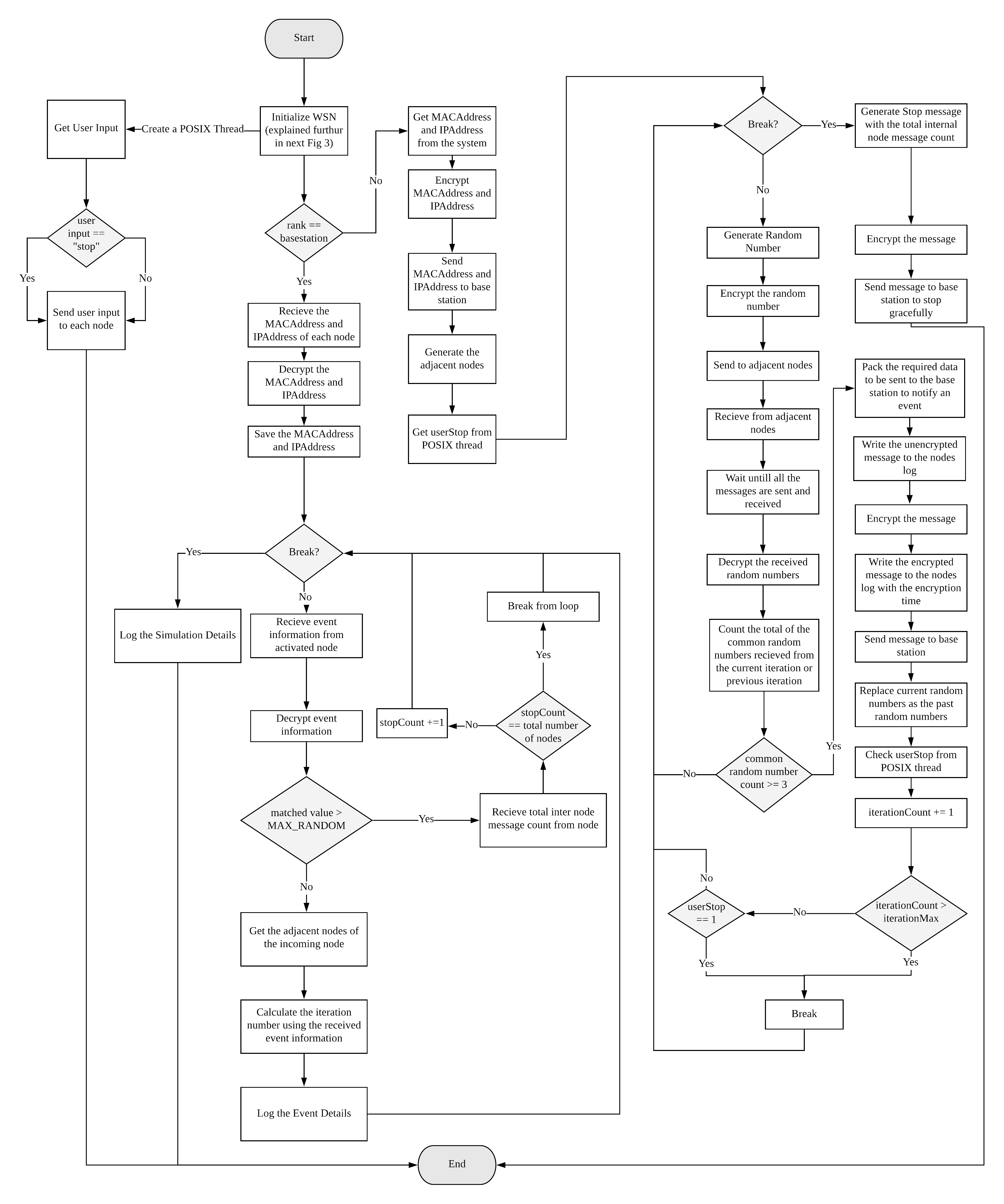}
		\caption{Technical Flowchart for WSN Inter Process Communication}
		\label{arc}
	\end{figure*}
		
	\begin{figure*}[!h]
		\centering
		\caption{Technical Flowchart for sub parts of the system}
		\subfloat[Technical Flowchart for Initializing the WSN]
		{
			\includegraphics[width=2.7in]{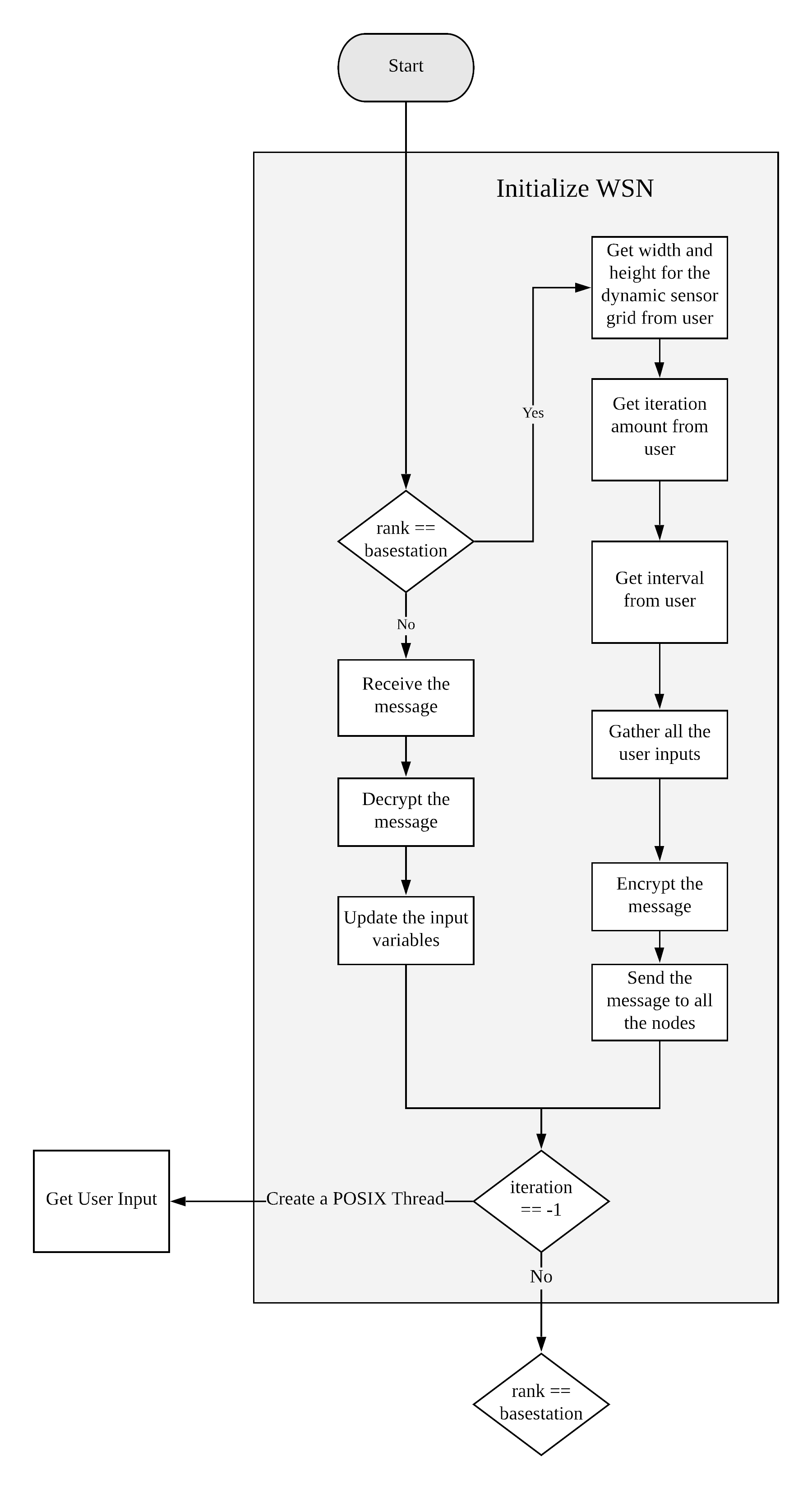}
			\label{inWSN}
		}
		\subfloat[Technical Flowchart for Encryption and Decryption algorithm with OpenMP]
		{
			
			\includegraphics[width=2.7in,keepaspectratio]{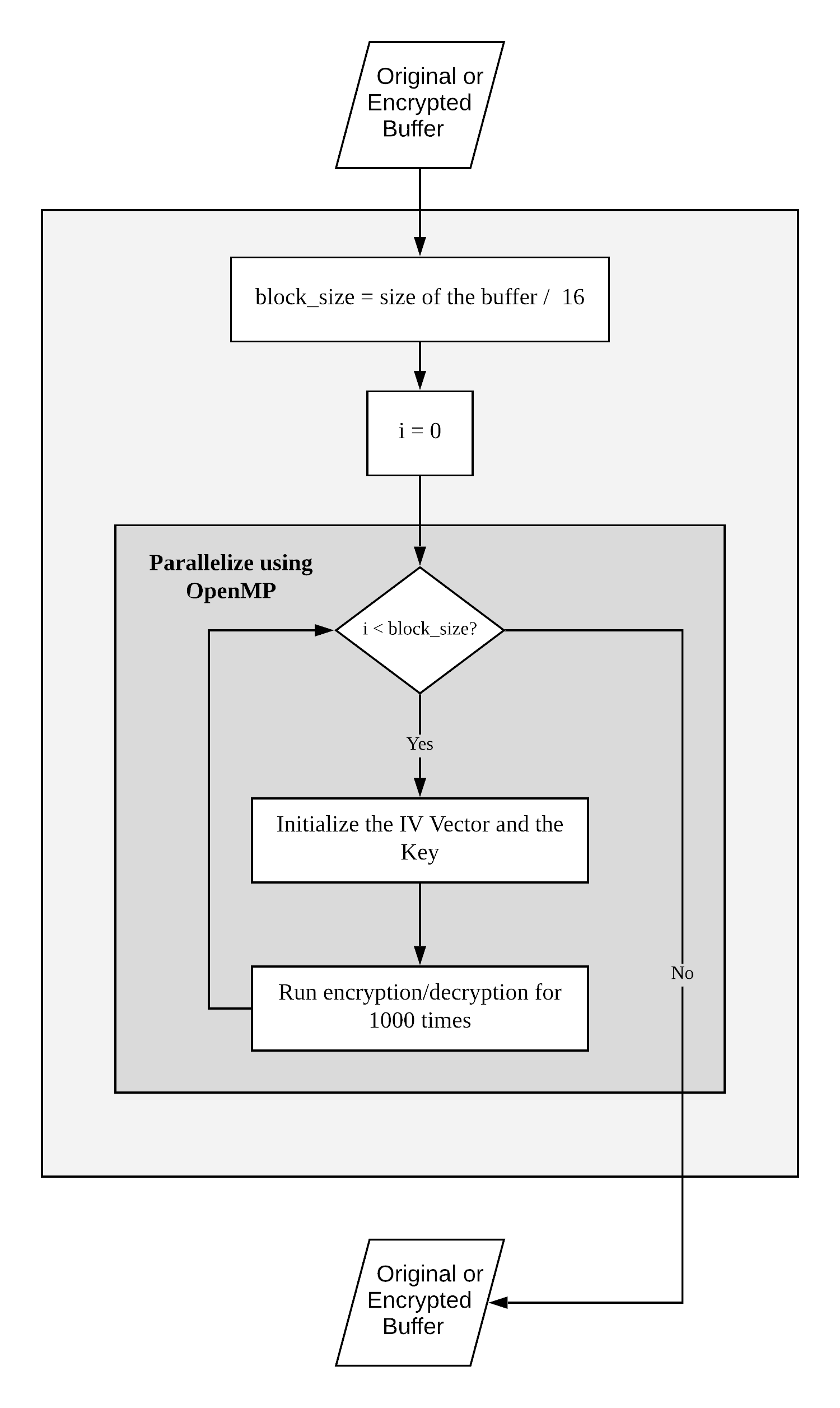}
			\label{encdec}
		}
	\end{figure*}

	\subsection{ Wireless Sensor Network Event Detection Algorithm }

	Now we look into the algorithm used by the system to detect events in the WSN. The system is divided into two main sections.
	
	\begin{itemize}
		\item Base Station which logs all the events
		\item Sensor Nodes which triggers and checks for events
	\end{itemize}

	Initially, the system generates the sensor network grid dynamically using the \emph{width} and \emph{height} given by the user which was discussed in Section \ref{IPCArc}. Then both the sensor nodes and the base station start running their respective methods. First the IP Address and the MAC Address is sent from each node to be stored in the base station for logging purposes. These messages are encrypted using the AES encryption algorithm \cite{daemen1999aes} discussed in section \ref{encryption}. The base station will then decrypt the message and store it in a dynamic array. Below is the data passed from the node to the base station when initializing the system.
	
		\begin{itemize}
		\item IP Address of the node
		\item MAC Address of the node
	\end{itemize}

	Figure \ref{arc} breaks down the algorithm of the whole system in to a technical flowchart. Then Figure \ref{inWSN} explains how the system initializes the WSN, based on the dynamic user inputs. Now let us dive deep into the algorithms used by the base station and the sensor nodes to generate and detect events.

	\subsubsection{Base Station}
	After the WSN is initialized, the base station starts checking for messages from any node. If the user has entered -1 as input for the iteration, the system will terminate checking for events only if the ``stop" keyword is entered by the user. In order for this to work, the base station creates a POSIX thread \cite{drepper2003native}, which enables it to check for user inputs as well as log the events from the nodes. When the user inputs the ``stop" keyword, the thread then sends a message to the nodes to stop event detection gracefully. Then the nodes will start sending the termination messages back to the base station and the system will shut down.
		
	\subsubsection{Sensor Nodes}
	
	Each node generates a random number and sends it to the adjacent nodes. This message which contains the random number, sent between the nodes are also encrypted. Then the sensor node starts receiving random numbers from the adjacent nodes. If three or more random numbers are the same, an event is detected. The algorithm also takes into account the random numbers generated by the nodes in the previous iteration. So it works as a sliding window with a time frame depth of 2. Then the nodes will notify the base station by sending a message. This message contains the following details:
	
		\begin{itemize}
		\item The random number that triggered the event
		\item An array of four elements which corresponds to the iteration of the matched value. If the left adjacent node matched with the random number of the previous iteration, the array will contain the iteration number of the previous element in the $0^{th}$ index of the array. 
		\item The MPI Wtime of the event detection to be used to calculate the communication time
		\item The date time string of the time, when the event was detected to be used for logging
	\end{itemize}
	
	After sending the message to the server, it saves the random numbers from the current iteration. If the maximum number of iterations is reached or the user's stop signal is received, the node will send a termination signal to the base station to exit out of the system gracefully.
	
	Non blocking send is used for the communication between the adjacent nodes, as blocking send has a chance of creating a deadlock. The communication between the sensor nodes and the base station is done using MPI blocking send and receive. This allows us to log the messages from the nodes as the data is being received compared to non blocking send.

	\subsection{ Encrypting communication and its effects }\label{encryption}
	
	Due to the increase risk of cyber attacks, all wireless networks needs to be encrypted as its is easier to attack and steal unencypted data. So, all the messages are encrypted throughout the wireless sensor network. After comparing multiple encryption standards, Advanced Encryption Standard (AES) was used to encrypt the messages \cite{daemen1999aes}.
	
	AES is a symmetric block cipher which can encrypt and decrypt information \cite{rouse_cobb}. The Encryption part of the algorithm converts data into cipher text form while decryption part converts cipher text into text form of data.  AES used different 128/192/256 bit keys to encrypt and decrypt data. For the WSN implementation we will be using AES 192, which uses a 192bit key. Since AES is the current standard for wireless communication, financial transactions and e-business we will be using it instead of other encryption algorithms \cite{rfwirelessworld}. As AES is a block cipher method, it has many modes of operations. Examples for these are Electronic Codebook (ECB), Cipher Block Chaining (CBC), Cipher Feedback (CFB) and Counter (CTR) \cite{meghna_2019}.
	
	For the WSN, we will be using the tiny-AES online implementation by kokke \cite{kokke_2019}, which use the Counter (CTR) block cipher mode. In order to improve the speed of the encryption process we will be using OpenMP \cite{openmp_2018} to parallelize the process. This implementation uses the same method to encrypt and decrypt the buffer, so the method takes in the original message or the encrypted message. Then the original / encrypted message is broken into 16 byte chunks. Then each chunk is encrypted 1000 times. We use OpenMP with dynamic scheduling to parallelize the outer loop. The reason for using dynamic scheduling is explored and analyzed in Section \ref{encry} of the report. The technical flow chart for the encryption and decryption algorithm is shown by Figure \ref{encdec}. The speed up analysis for the encryption and decryption algorithm is discussed in Section \ref{encry}.
	
	\section{Methodology}\label{method}

	In this section we look into the methodology used to test and analyze both the event detection criteria as well as the encryption algorithm used in the WSN. Table \ref{params} summaries the parameters used by the program to get the data required for analysis. In order to make the encryption and decryption time consistent, the buffer size for all the communication is constant. So the single random number communication between the sensor nodes uses this buffer size. This is inefficient, but allows us to accurately record the encryption and decryption times for all the messages in the network. 
	
	In order to analyze the encryption speed up, the three simulations are executed. One without using OpenMP, and the other two with static and dynamic scheduling of OpenMP \cite{jakas_corner_2016}.
	
	All of the test cases were run on an Intel® Xeon® W-2145 Processor, which contains 8 cores and 16 logical threads. Figure \ref{screen 1} in the Appendix shows the screen shot of the console while testing. Another approach for the test case would be to use the dynamic stopping approach. Figure \ref{screen 2} in the Appendix shows the screen shot of the console for this approach. 
	
	For testing purposes, the time taken for encryption is logged by each node into a csv file and the other required data is also logged into a csv file from the base station.
		
	\begin{table}[!h]\caption{Specifications of the system and the parameters used for generating the data to be analyzed}
		\begin{center}
			\renewcommand{\arraystretch}{1.2}
			\begin{tabular}{| c | c | c |} 
				\hline
				\textbf{Specification} & \textbf{Value} & \textbf{Description} \\ \hline
				\emph{CPU}    &  8/16   & \makecell{Number of CPUs and \\ logical cores }   \\ \hline
				\emph{RAM}    &  64   & \makecell{Memory of the system \\}   \\ \hline
				\textbf{Parameter} & \textbf{Value} & \textbf{Description} \\ \hline
				\emph{Width}    &  4   & \makecell{The width of the sensor \\ node grid }   \\ \hline
				\emph{Height}    &  5   & \makecell{The height of the sensor \\ node grid }   \\ \hline
				\emph{Iterations}    &  100   & \makecell{Total number of iteration \\ performed by the system }   \\ \hline
				\emph{Interval}    &  1   & \makecell{Wait time between \\ each iteration }   \\ \hline
				\emph{MaxRandom}    &  12   & \makecell{Limit the random numbers \\ generated to between 0 - 11}   \\ \hline
				\emph{Packsize}    &  256  & \makecell{Size of the buffer for sending\\ and receiving data }   \\ \hline
			\end{tabular}
			\label{params}
		\end{center}
	\end{table}

	\section{Results and Discussion} \label{res}
	
	This section analyzes the results obtained using the methodology described in Section \ref{method}. The analysis is broken down into three main sub sections, each explaining the different aspect of the result. First the structure of the log files are explained and the data of the log files are analyzed. Then the communication time is analyzed. Finally the encryption algorithm and the speed up of using OpenMP is analyzed.
	
	\subsection{Summary of Events} \label{eventsum}
	
	When an event is triggered by the sensor, the system generates two types of log files. 

	\begin{itemize}
		\item Base Station generates the main log file which logs the event information and the simulation information
		\item Each sensor node logs the the original message, encrypted message and the encryption time
	\end{itemize}

	For readability purpose, parts of the log files will only be displayed in the report. The full log files and the testing files are included with the submission for the validity of the tests.

	\subsubsection{Base Station Log File}
	
	The base station logs two types of information, the event detection details and the simulation details at the end. First lets look into the log files to understand each part of it. 
	
	\begin{figure}[!h]
		\centering
		\includegraphics[width=3.3in,keepaspectratio]{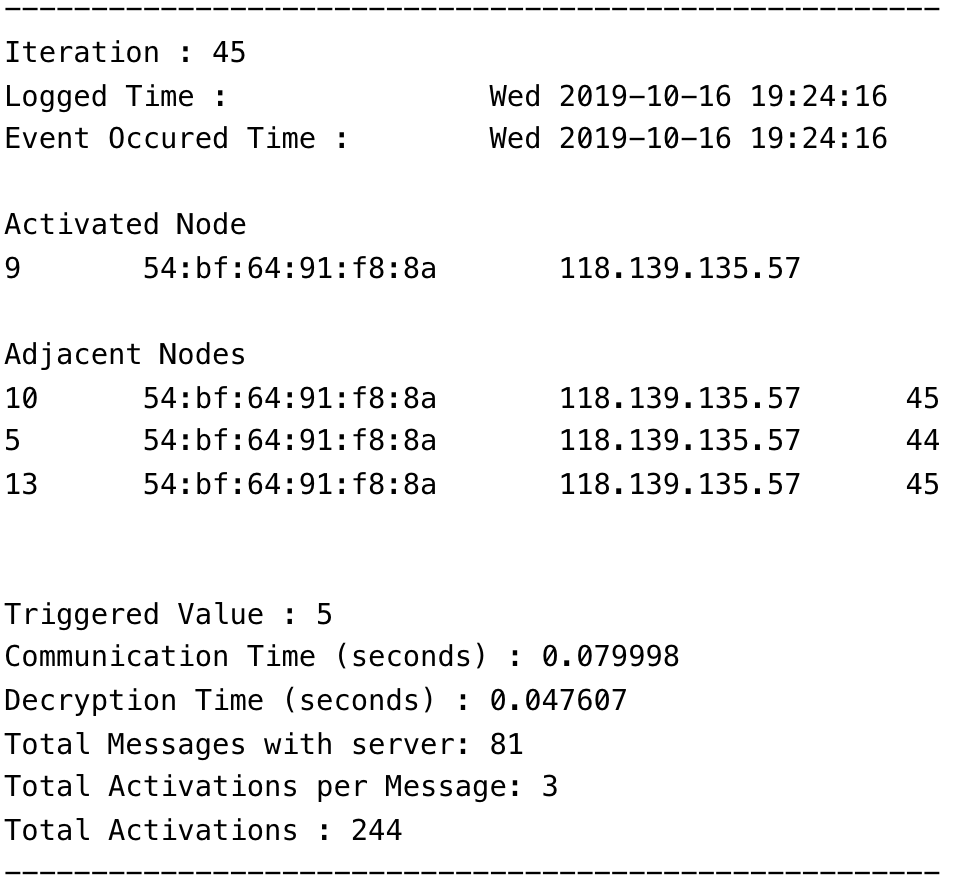}
		\caption{Base station log file for event details}
		\label{lst:log1}
	\end{figure}	
	Figure \ref{lst:log1} shows an event triggered at the 45 iteration of a simulation. This can be observed from the first line of the log which determines the iteration number. The next two lines of the log displays the time when the event was logged on the file and the actual time when the event is detected respectively. This is useful when the node and base station is present in two geographical locations and the time to send the message effects the logging time. But in this example, since the node is a process inside the same machine, we are not able to see a difference. The next two lines denote the activated node that was triggered i.e. the node that sent the message to the base station. The Rank, MAC Address and IP Address of the activated node is being logged in the log file. In a case where each node were running on separate machines, we are able to see different MAC and IP addresses. Next the adjacent nodes which generated the same random number is being logged. Similar to the activated node the Rank, MAC Address and IP Address of the adjacent nodes is also being logged in the log file. But there is also a fourth item that stores the iteration number. Since the WSN takes into account the random values generated from the previous iteration, this value helps to identify in which iteration the same random number was generated.
	
	The next part of the log file shows the random number that triggered the event. Then the communication time and the decryption time is logged. The total messages with the base station shows how many messages are being sent to the server up until the current iteration. The total activations per message shows how many nodes are triggered for the current message. This can be either 3 or 4. Then finally the total activations up until the current iteration is also logged.

\begin{figure}[!h]
		\centering
		\includegraphics[width=3.3in,keepaspectratio]{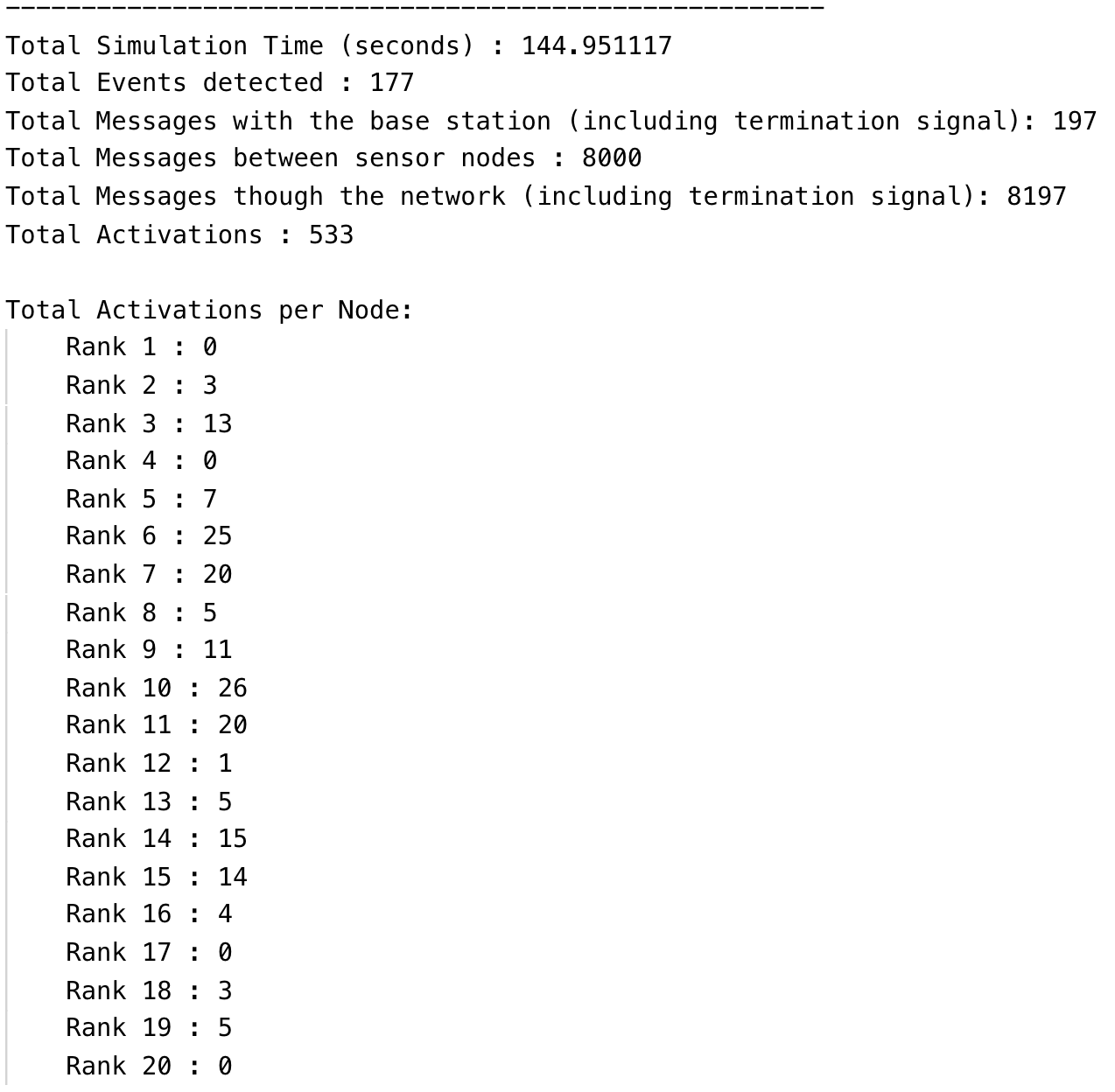}
		\caption{Base station log file for simulation details}
		\label{lst:log2}
	\end{figure}

	Figure \ref{lst:log2} displays the log file that is generated at the end, when the system shuts down gracefully. It records the total simulation time, total number of events detected, total messages to the base station, total sensor node to sensor node messages and the total messages through the network. The total messages through the network contains the messages sent between the sensor nodes, messages sent from node to base station when an event is triggered and finally the messages sent from the  sensor nodes with the termination signal to stop gracefully. It then logs the total activations through out the network and how many times each node was activated during the simulation  according to the rank. 
	
	\subsubsection{Sensor Node Log File}
	
\begin{figure}[!h]
		\centering
		\includegraphics[width=3in,keepaspectratio]{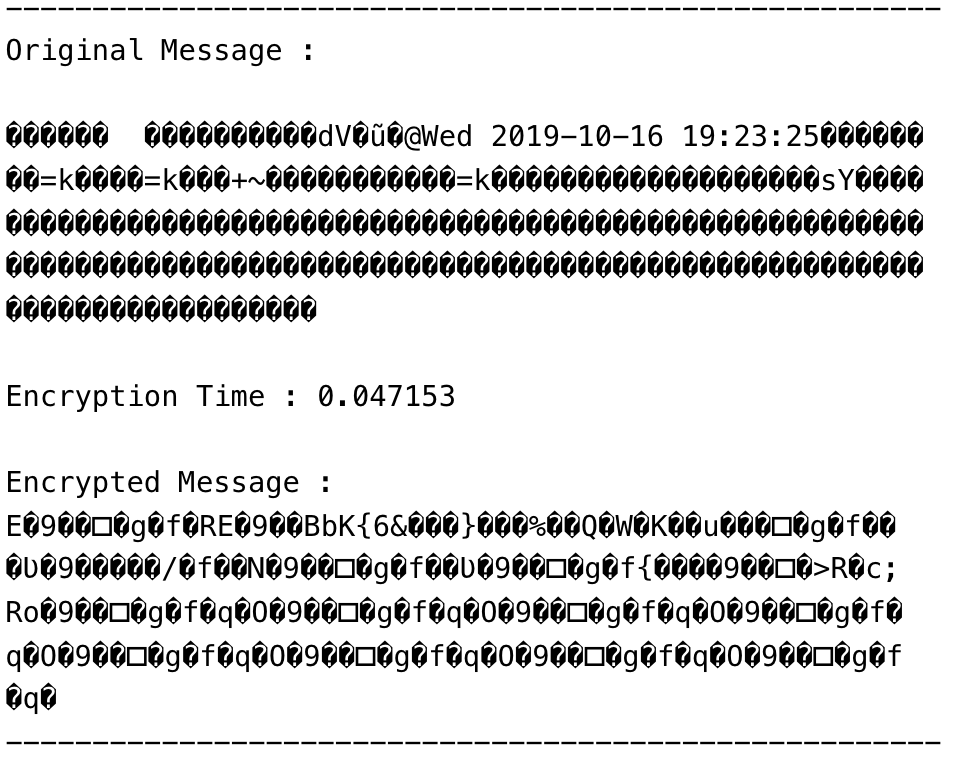}
		\caption{Sensor Node log file}
		\label{Message}
	\end{figure}

	Next we look into the log file of each sensor node. Figure \ref{Message} shows the first event of the log file generated by the node with rank 3. This log file contains the original un-encrypted message, the encryption time and the encrypted message. If we analyze the log file we can see the time stamp in the original message but, after encrypting we do not see it.
	
	\subsection{Inter Process Communication Analysis} \label{comm}
	
	In this section we analyze the WSN data collected from the simulation.

	\subsubsection{Event detection messages from sensor nodes to base station} \label{ob}
	
		\begin{figure*}[!h]
		\centering
		\caption{Analysis of the data generated from the WSN}
		\subfloat[Total Messages per Iteration]
		{
			\includegraphics[width=6in]{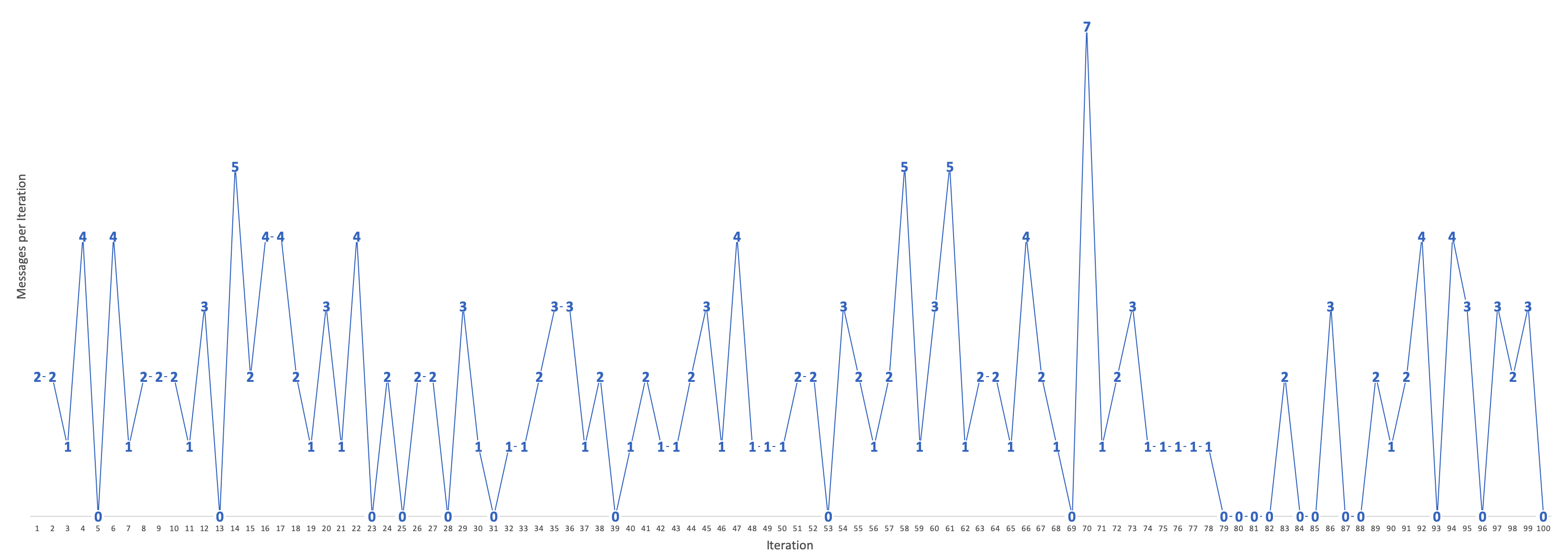}
			\label{me}
		}
		\\
		\subfloat[Average Communication Time per Iteration]
		{
			
			\includegraphics[width=6in]{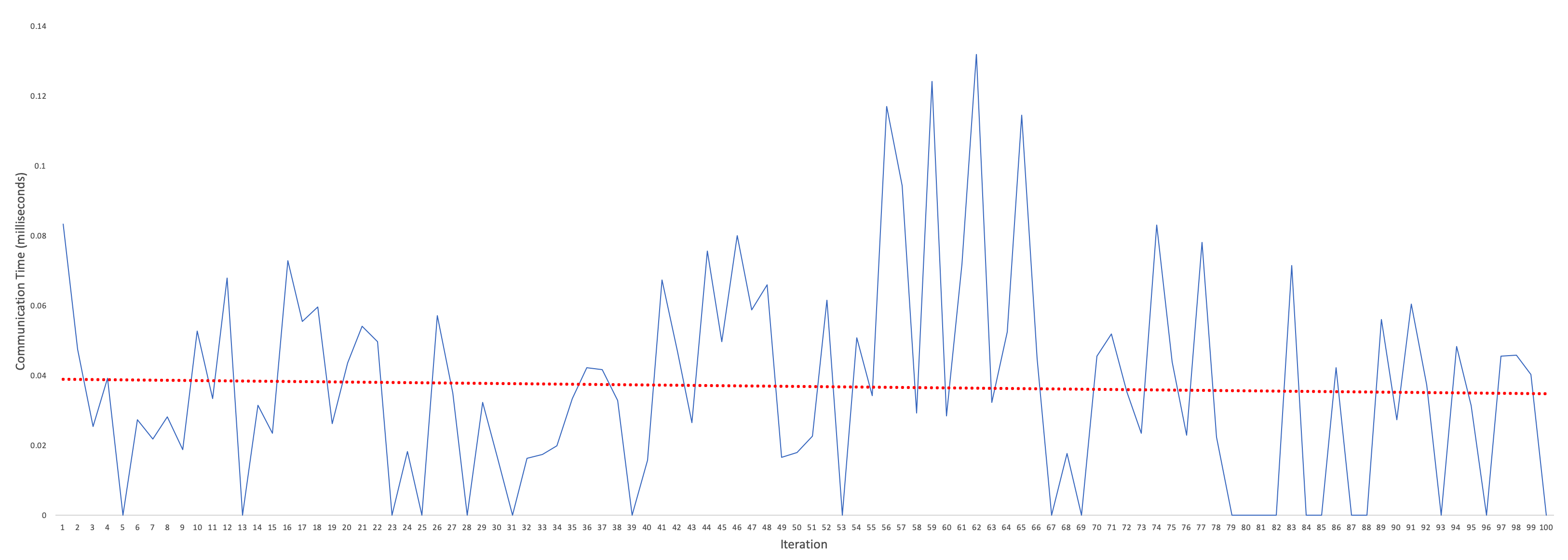}
			\label{com}
		}
	\end{figure*}	

	 For each event detection, this architecture sends one message from the activated node to the base station. So referring to Figure \ref{lst:log2}, the total number of events generated is 177 and the total number of messages received from the nodes to the server is 197 which includes the 20 termination signals sent at the end. Figure \ref{me} plots the number of messages received from the nodes over iterations. 7 out of the 20 nodes were activated at iteration 70, which is the highest amount of nodes activated per iteration. We can also see that after every iteration that has a high number of messages, the number of messages plummets down. The reason for this is, commonly high number of messages occur when there are matching random numbers in both the current iteration and the previous iteration. So after a high message iteration the previous array mostly contains 2 or more of the same common random numbers. So the probability of the random number of the current iteration being the same is low. Therefore the number of events detected in the next iteration goes down.
	 
	 \begin{figure}[!h]
	 	\centering
	 	\includegraphics[width=3.7in,keepaspectratio]{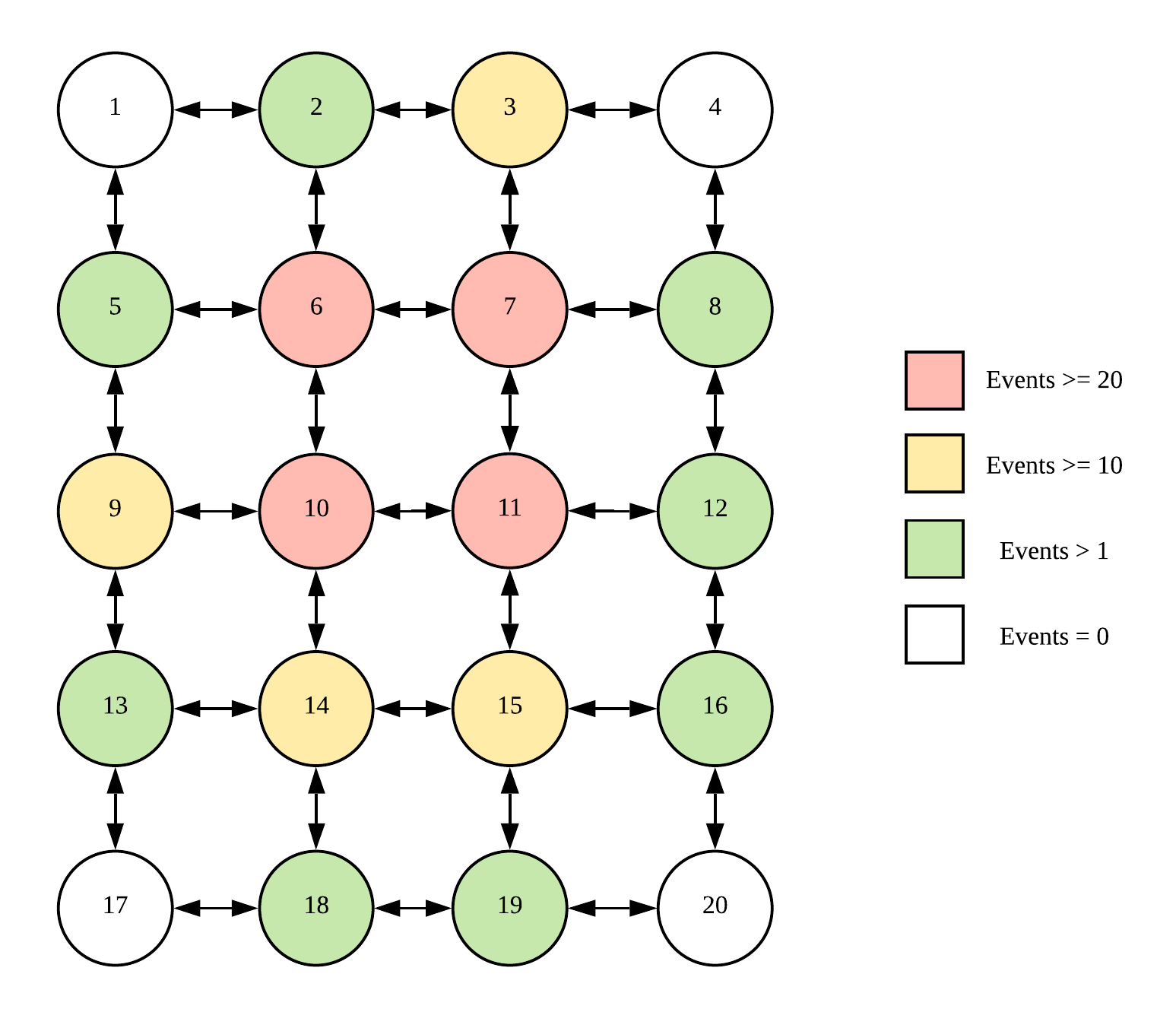}
	 	\caption{Sensor Grid overlayed with the total event detection from each node}
	 	\label{colgrid}
	 \end{figure}
 
	\subsubsection{Total Events detected from each node}
	Next lets look at Figure \ref{colgrid}, which is a heat map of the nodes and the events detected from each of them. The sensor nodes with four adjacent neighbors have the highest number of events generated as it has more probability of having 3 or more common random numbers. Also note the four corner nodes never detect an event as they only have 2 adjacent nodes. Therefore, the observations satisfy the actual probability of the event detection where the center nodes detect more events.
	
	\subsubsection{Communication time}
	
	The communication time includes the time for encrypting the message, sending the message, receiving the message and decrypting the message. Since the simulations are run on a single computer, the communication time from node to node must be similar in theory. But using Figure \ref{com} we see that the times are different. The reason for this can be obvious if we compare Figure \ref{me} and \ref{com}. As a high number of events  are detected close to each other, the communication time increases. This is due to the blocking communication used between the nodes and base station. But the red line in Figure \ref{me} shows the tread line of the communication time. So even though there are peaks in the average communication time, the trends shows a constant speed as is should be in theory. Therefore, we can confirm that our assumption of constant time is true.
	
	Another issue we might face is that when one part of the network is in a separate geographical location to the base station, the communication time will increase. So the nodes with high communication time will take longer to finish the iterations. So the base station will wait until all the nodes have sent the termination signal to avoid any issue. 

	\subsection{Encryption Analysis}	\label{encry}
	
	In this section we will analyze how OpenMP helps improve encryption and decryption time of the AES algorithm.
	
	In order to compare the speed up, first the encryption and decryption time of the AES algorithm without OpenMP is recorded. The encryption time is extracted from the node log files, where as the decryption time is extracted from the base station log file. 
	
	OpenMP has a static and dynamic way to parallelize a for loop \cite{jakas_corner_2016}. In order to determine which way is better, simulations are run for each method and the speed is is calculated. This is done for both the encryption and decryption process. 
	
	We can calculate three types of speed up in the WSN. 
	
	\textit{Note : $AT$ represents the average time, while $SP$ represents the speed up.}
	
	\begin{itemize}
		\item Speed up of Encryption which can be obtained by Equation \ref{spen}.
		
		\begin{equation}
		SP_{encryption} = \frac{AT_{encryption}^{serial}}{AT_{encryption}^{parallel}}
		\label{spen}
		\end{equation}
		
		\item Speed up of Decryption which can be obtained by Equation \ref{spdc}.
		
			\begin{equation}
		SP_{decryption} = \frac{AT_{decryption}^{serial}}{AT_{decryption}^{parallel}}
		\label{spdc}
		\end{equation}

		\item Speed up of Encryption and Decryption which can be obtained by Equation \ref{spt}. But in order to calculate this speed up we first get the average encryption time and decryption time, which is denoted by Equation \ref{avgt}.
		
		\begin{equation}
		AT_{total} = \frac{AT_{encryption} + AT_{decryption}}{2}
		\label{avgt}
		\end{equation}
		
		\begin{equation}
		SP_{total} = \frac{AT_{total}^{serial}}{AT_{total}^{parallel}}
		\label{spt}
		\end{equation}
		
	\end{itemize}
	
\begin{figure}[!h]
		\centering
		\includegraphics[width=3in,keepaspectratio]{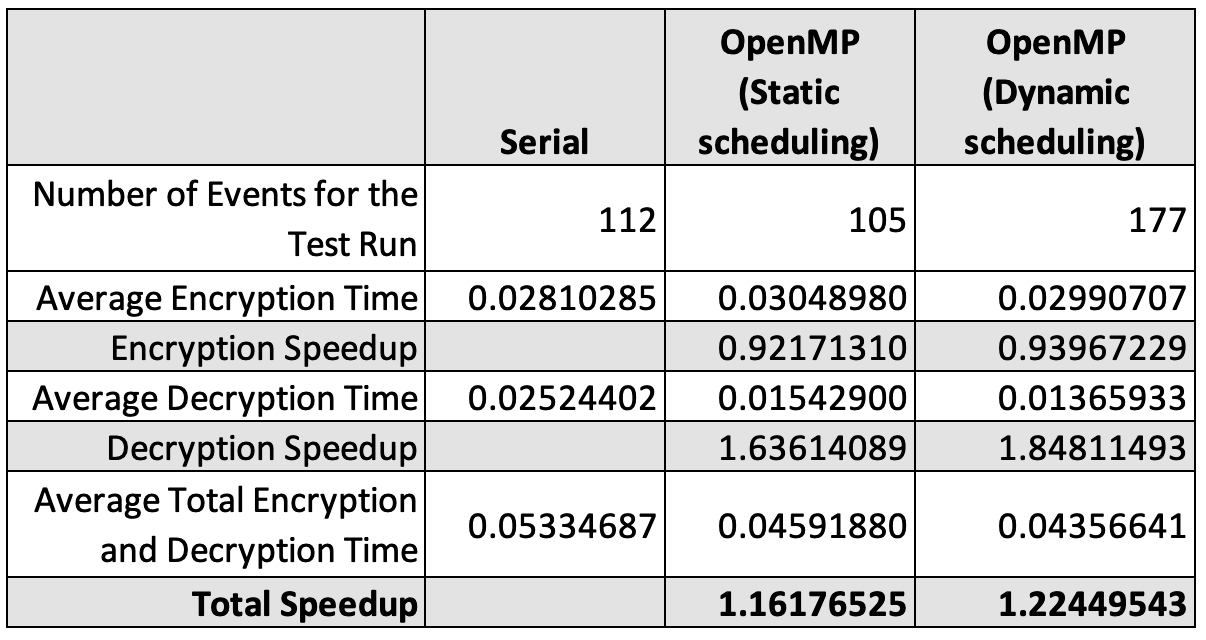}
		\caption{Analysis of the OpenMP speed up for encryption and decryption}
		\label{speedup}
	\end{figure}

	Figure \ref{speedup}, shows the results obtained by this experiment. We will now analyze the speed up based on two criteria.
		
	\subsubsection{Encryption, Decryption and Total Speed up}
	  If we analyze the results we see that the speed up for encryption time is lower than 1, but the decryption time and the total speed up is more than 1. The reason for this is because we are using a single computer for running the WSN. So each time multiple events are detected through out the network, the threads that are used for encryption by each node, floods the system. So on average the speed will be lower. but if we consider decryption, we can see a better speed up, as only one node i.e. base station creates the threads. But if we consider the total speed up for encryption, which includes both encryption and decryption, the speed up is higher than 1. This can be improved further by running each sensor node on different processor as it will have free threads to improve the time.
	  
	  \subsubsection{Static vs Dynamic Scheduling}
	  
	  Looking at the results, dynamic scheduling is better than static scheduling. The main reason for this that some threads get prioritized by the system scheduler. So if its dynamic, the high priority threads are able to perform more cycles compared to the static approach, which assigns the low priority threads with the same workload. Therefore dynamic scheduling is more suitable for parallelizing the encryption and decryption process.

	\section{Conclusion}
	
	Taking into account the hypothesis made in Section \ref{intro}, and all the observations and results made through the simulations in Section \ref{res}, let us condense the core outcomes of this report. 
	
	Section \ref{theoratical} delved into the theoretical and algorithmic details of the grid based IPC architecture. A high level technical explanation of the AES encryption algorithm and the benefits of it was also discussed. 
	
	Then Section \ref{method} proposed a testing methodology to observe the effects of the proposed architecture. Three main criteria were taken into account when discussing the results of the simulations. Section \ref{eventsum} summarized the logging portion of the system and then Section \ref{comm} and \ref{encry} analyzed how the communication time and encryption time affects the overall performance in a grid based architecture.
	
	Now let us epitomize the two main finding of this report. Firstly the communication time of a grid based architecture can be dependent on the number of events as the base station causes a bottle neck when writing to the log files in serial. Next the encryption algorithm brings in more security to the system with the cost of more power and time to send events from the sensor nodes to the base station. This can be improved by using OpenMP to implement thread level parallelism, but, this in turn will bring in more power and resource usage which are limited in a WSN.
	
	Conclusively, we can state that the grid based architecture has the benefit of faster event detection, as there is a direct link from the sensor nodes to the base station. This in turn allows the sensors to fire only when there is an event, so they can be in a power efficient state. We can also conclude that the parallelized AES encryption algorithm is able to reduce the communication times and also provide more security for the wireless sensor networks.

	\section{Future Work}
	
	Many future research paths are opened as a result of this report. One of the possible paths to explore is the effects of sensors nodes in multiple geographical locations and how it effects the communication time. The fixed buffer size assumption that was taken in order to observe the speed up of the encryption algorithm can be removed and the affects of dynamic message length can also be explored. Then the observation that was made in Section \ref{ob} where the messages plummet down after a set of high activation messages can be proved formally by testing it in a controlled environment. Finally, the system can be implemented in a prototype environment to observe the energy graphs of each node and analyze how the events and encryption effects the power consumption of the system.
	
	Therefore, considering the aforementioned proposed future paths, we can state that this report paves the way to many exciting research areas on how an efficient Grid based Inter Process Communication Architecture affects Wireless Sensor Networks.
	
\bibliographystyle{IEEEtran}
\bibliography{ref.bib}

	\newpage
	\onecolumn
	\appendix

\begin{figure*}[h]
	\centering
	\includegraphics[width=6in,keepaspectratio]{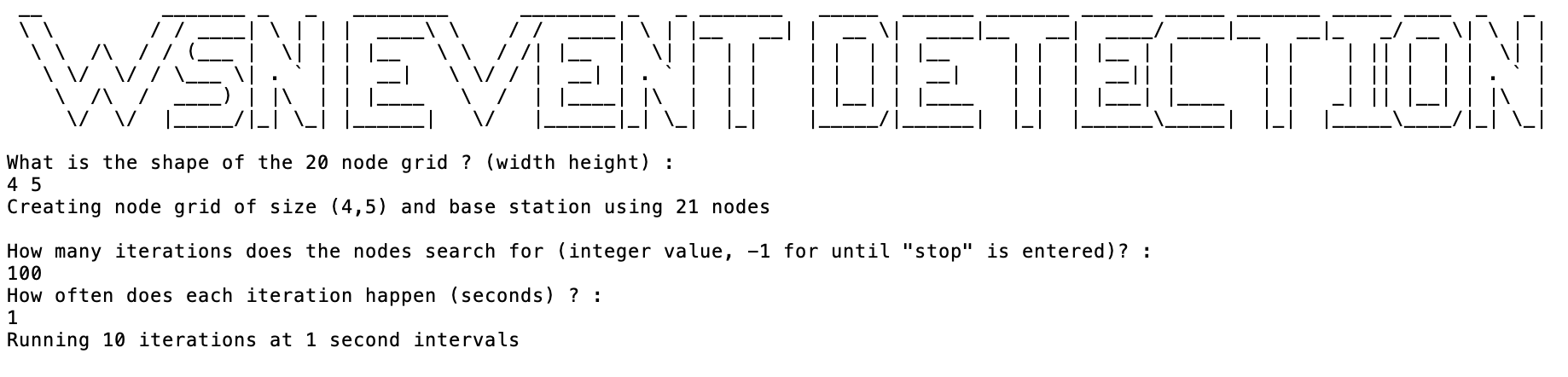}
	\caption{Screenshot of the console used for generating of the results}
	\label{screen 1}
\end{figure*}

\begin{figure*}[h]
	\centering
	\includegraphics[width=6in,keepaspectratio]{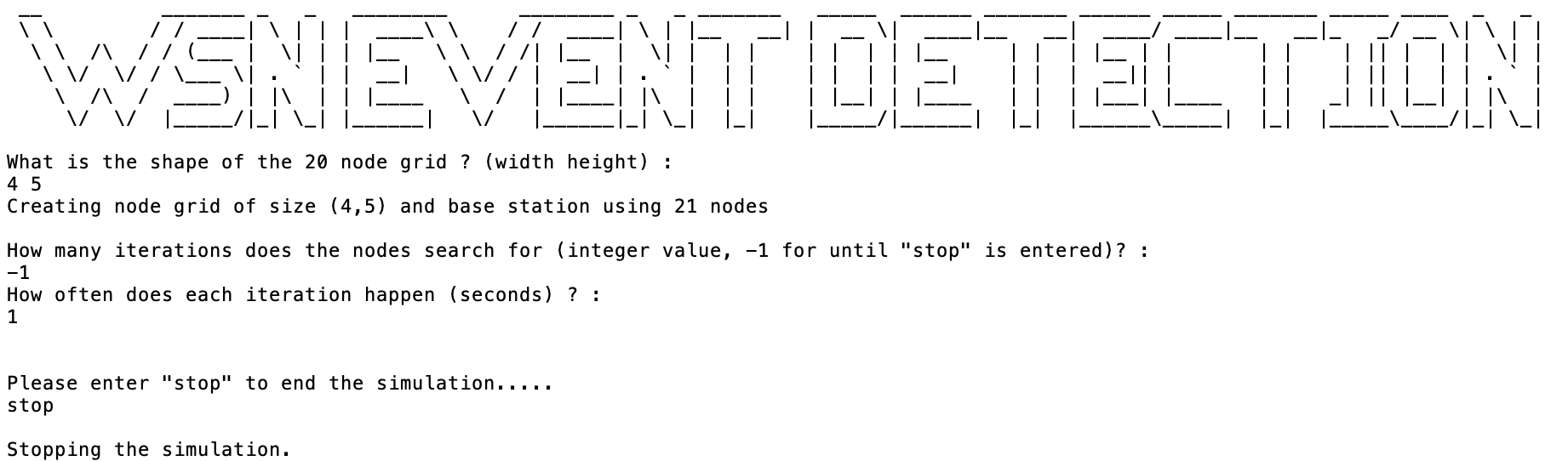}
	\caption{Screenshot of the alternative approach to run the program until user enters stop}
	\label{screen 2}
\end{figure*}

\end{document}